\documentclass[12pt]{article}
\usepackage{url}
\usepackage{graphicx}
\usepackage{float}
\usepackage{sansmath}
\usepackage{xcolor}
\usepackage{setspace}
\usepackage{amsmath}
\usepackage{natbib}
\usepackage{authblk}
\usepackage{lipsum}
\usepackage[section]{placeins}
\providecommand{\keywords}[1]{\textbf{\textit{Keywords:}} #1}

\begin{document}
\title{A Probabilistic Model for Predicting Shot Success in Football}
\author{Edward Wheatcroft and Ewelina Sienkiewicz}
\affil{\footnotesize London School of Economics and Political Science, Houghton Street, London, United Kingdom, WC2A 2AE.}
\affil[]{\footnotesize Corresponding email: e.d.wheatcroft@lse.ac.uk}
\date{}
\maketitle
\noindent

\begin{abstract}
Football forecasting models traditionally rate teams on past match results, that is based on the number of goals scored. Goals, however, involve a high element of chance and thus past results often do not reflect the performances of the teams.  In recent years, it has become increasingly clear that accounting for other match events such as shots at goal can provide a better indication of the relative strengths of two teams than the number of goals scored.  Forecast models based on this information have been shown to be successful in outperforming those based purely on match results.  A notable weakness, however, is that this approach does not take into account differences in the probability of shot success among teams. A team that is more likely to score from a shot will need fewer shots to win a match, on average. In this paper, we propose a simple parametric model to predict the probability of a team scoring, given it has taken a shot at goal. We show that the resulting forecasts are able to outperform a model assuming an equal probability of shot success among all teams. We then show that the model can be combined with predictions of the number of shots achieved by each team, and can increase the skill of forecasts of both the match outcome and of whether the total number of goals in a match will exceed 2.5.  We assess the performance of the forecasts alongside two betting strategies and find mixed evidence for improved performance. 
\end{abstract}

\keywords{Probability forecasting, Sports forecasting, Football forecasting, Expected goals, Shot success}

\section{Introduction}
Association football (hereafter football) is by far the most popular sport globally with almost every country in the world having a national team and often a multitude of domestic leagues.  In England alone, there are over 100 professional teams.  The vast popularity of the sport has led to demand for predictive information regarding the outcomes of matches, competitions and leagues, often driven by the desire to gamble on them.  In recent years, a vast number of betting markets have opened up, providing opportunities for those with useful predictive information and/or insight to make a profit. Betting strategies are usually underpinned with predictive models that attempt to predict the probability of different outcomes, thus informing which bets to take. \par 

Whilst early attempts at building predictive models focused on the number of goals scored by each team in previous matches, in recent years it has become clear that there is predictive value in other match events such as the number and nature of shots and corners taken by each team (\cite{wheatcroft2019evaluating,wheatcroft2020}).  The key insight is that the number of goals scored by each team is subject to a higher level of chance than events such as shots and corners, which can be more reflective of the quality of the performances of the teams. Take, for example, a match in which the home team takes a large number of shots but is unable to score, whilst the away team takes few shots and happens to score from one of them and win the match. A predictive model that only takes goals into account would not reflect the fact that the home team dominated the match and may wrongly downgrade the forecast probability that they win future matches. \par

In this paper, we consider a large number of football matches in which match data such as the number of shots, shots on target and corners is provided.  In order to build a set of match forecasts, we are then interested in (i) the number of shots taken by each team and, (ii) the probability that any shot results in a goal. Given these two ingredients, we can then predict the number of goals scored by each team in a match. For (i), we make use of a recently developed methodology which uses a rating system to predict the number of shots taken by each team.  For (ii), we propose a simple model to predict the probability of scoring from a shot. The latter is tested on over one million shots from European football matches in 22 different leagues and, when calibrated, is shown to be capable of producing skillful probabilistic forecasts.  Forecasts of the number of shots are then combined with forecasts for the probability of shot success to construct forecasts of both the match outcome and whether the total number of goals in a match will exceed $2.5$. \par

The focus of this paper is on assessing the probability of a team scoring conditioned on them taking a shot at goal. In fact, the question of how to assess the probability of scoring from a shot is one that has received a lot of attention in the football forecasting literature.  However, the focus has almost exclusively been on factors such as the location and nature of the shot, position of players etc.  Here, we do not attempt to take this information into account and rather estimate the probability of shot success on past data, focusing on the strengths of the teams.  This is not simply a limitation of our methodology but a property of the question we are trying to address. We look to estimate the probability of shot success before the match has started and therefore we cannot condition on the specific nature of each shot.  Whilst we can attempt to predict the number of shots taken by each team, it is not realistic to be able to predict the nature of those shots. The output of our model is therefore a fixed forecast probability of shot success for each team in a match. \par

Typically, the nature of football prediction models is that each team involved in a league or cup competition is given a `rating'.  These rating systems often take one of two different approaches.  In the first, each team's rating is a variable which is updated as new information emerges. The nature of those updates are governed by a small number of parameters which determine aspects such as the effect of the result of the last match or the margin of victory/defeat.  We refer to these as \emph{Variable Rating Systems}.  The other category assigns each team one or more parameters which determine their strength and these are usually estimated using maximum likelihood (\cite{ley2019ranking}). In that case, a large number of parameters are required to be estimated simultaneously and fairly sophisticated optimisation algorithms are often needed.  We deviate from the terminology used by \cite{ley2019ranking}, who refer to such models as `Maximum Likelihood models' and, instead, use the more general term \emph{Parametric Rating Systems}.  In this paper, we make use of both approaches.  Our shot probability model (the novel model in this research) is a Parametric Rating System which assigns attacking and defensive ratings to each team and these are estimated using maximum likelihood estimation.  In addition, we make use of a Variable rating system in the form of Generalised Attacking Performance (GAP) ratings which estimate the number of shots achieved by each team (\cite{wheatcroft2020}). \par

There is a large body of literature proposing approaches to building ratings systems for sports teams or players.  By far the most well known approach is the Elo rating system which has a long history in sport and has inspired many other systems.  Elo ratings were initially designed with the intention of providing rankings for chess players and the system was implemented by the United States Chess Federation in 1960 (\cite{elo1978rating}).  The Elo system assigns ratings to each player or team, which are then used to estimate probabilities of the outcome of a game.  The rating of each player is then updated to take the result of the game into account. Whilst the system was initially designed for cases in which the outcomes are binary (i.e. there are no ties), more recently, it has been extended to account for draws so that they are applicable to sports such as football, in which draws are common.  After each match, the system takes the difference between the estimated probabilities and the outcome (assigned a one, a zero, or $0.5$ for a draw) and adjusts the ratings accordingly. The system in its original form therefore does not account for the \emph{size} of a win.  Elo ratings have been demonstrated in the context of football and shown to perform favourably with respect to six other rating systems (\cite{hvattum2010using}). FIFA switched to an Elo rating system in 2018 to produce its international football world rankings (\cite{Fifa_elo}).  Elo ratings are also common in other sports such as Rugby League (\cite{carbone2016rugby}),  American Football (\cite{538}) and Basketball (\cite{538NBA}). \par

Whilst Elo ratings have been an important part of sports prediction for many years, they are limited in that they do not directly take home advantage into account.  This is important because home advantage has a very big effect in football (\cite{pollard2008home}).  Adjustments have been made to the Elo rating system to account for this but this typically consists of a single parameter that doesn't account for variation in the home advantage of different teams (\cite{538NBA,538}). Rating systems such as the GAP rating system used in this paper distinguish between home and away performances by giving separate ratings for each.  This is also true of the pi-rating system introduced by \cite{constantinou2013determining}, for example.  \par
 
Variable Ratings Systems such as the GAP rating system assign ratings to each team which are updated each time they are involved in a match. Similar approaches have been taken by a large number of authors. For example, \cite{maher1982modelling} assigned fixed ratings (i.e. not time varying) to each team and used them in combination with a Poisson model in order to estimate the number of goals scored.  A similar approach was used by \cite{dixon1997modelling} to estimate match probabilities.  It was shown that the forecasts were able to make a statistically significant profit for matches in which there was a large discrepancy between the estimated probabilities and the probabilities implied by the odds. The Dixon and Coles model was modified by \cite{dixon2004value} who were able to demonstrate a profit using a wider range of published bookmaker odds. A Bayesian model which produced time-varying attacking and defensive ratings was defined by \cite{doi:10.1111/1467-9884.00243}.  There are many other examples of systems that use attacking and defensive ratings and these can be found in, for example, \cite{karlis2003analysis}, \cite{lee1997modeling} and \cite{baker2015time}. 

A number of authors have taken a Parametric Rating System approach to modelling football matches.  An overview can be found in \cite{ley2019ranking} in which a Bivariate Poisson model is shown to produce the most favourable results according to the Ranked Probability Score (RPS). A profitable betting strategy has also been demonstrated by \cite{koopman2015dynamic} using a Bivariate Poisson model. The approach taken by \cite{ley2019ranking}, in which less recent matches are weighted lower than more recent matches, provides inspiration for our shot success model. \par

Related to the prediction of shot success is the concept of `expected goals' which has been growing significantly in prominence in football analysis in recent years. The rationale is that the nature of a team's attempts at goal can be used to estimate the number of goals they would be `expected to score' in a match. For a particular shot, the `expected' number of goals is simply the estimated probability of scoring given characteristics such as the location, angle to goal, position of defenders etc.  As a result, a great deal of effort has been made to model the probability of scoring based on information of this kind. For example, \cite{ruiz2015measuring} attempt to evaluate the efficacy of football teams in terms of converting shots into goals by taking account of characteristics such as the location and type of shot (e.g. whether the shot was taken from open play). \cite{gelade2014evaluating} built a model to evaluate the performance of goalkeepers by taking the factors such as the location, deflections and swerve of the ball into account.  Many other papers have been written on the subject and a good overview can be found in \cite{eggels2016expected} and \cite{rathke2017examination} who also present their own models. \par

The main aim of this paper is to define and demonstrate a model for the probability of a team scoring from a shot in a football match. To our knowledge, whilst significant effort has been made to estimate probabilities of scoring given the specific nature of a shot (such as location), none of these approaches attempt to provide predictions of shot success before the match and cannot be used for this purpose.  In short, the aim of those models is to predict the probability of scoring from a particular shot given various characteristics, whilst the purpose of our model is to predict the probability of scoring given the strengths of the teams involved and the location of the match (i.e. which team is at home).  The latter can easily be combined with predictions of the number of shots achieved to predict the overall number of goals for each team. \par

This paper is organised as follows.  In section~\ref{section:data}, we describe the data set used to demonstrate our model.  In section~\ref{section:model}, we describe our model of shot success and assess its performance in terms of forecast skill and reliability in 22 different football leagues.  In section~\ref{section:predicting_match_outcomes}, we demonstrate the use of our shot success model in combination with the GAP rating system to provide forecasts of match outcomes and whether the total number of goals in a match will exceed 2.5.  Section~\ref{section:discussion} is used for discussion.

\section{Data} \label{section:data}
In this paper, we make use of the football data repository available at \url{www.football-data.co.uk}, which supplies match-by-match data for 22 European Leagues. For each match, a variety of statistics are provided including the number of shots, shots on target and corners.  In addition, odds data from multiple bookmakers are provided for the match outcome market, the over/under 2.5 goal market and the Asian Handicap match outcome market.  For some leagues, match statistics are available from the 2000/2001 season onwards whilst, in others, these are available for later seasons only.  Since we require shot data, only matches from the 2000/2001 season onwards are considered.  A summary of the data used in this paper is shown in table~\ref{table:Leagues_available}.  Here, the total number of matches since 2000/2001, the number of matches in which shots and corner data are available and the number of these excluding a `burn-in' period for each season are shown. The `burn-in' period is simply the first six matches of the season for the each team.  This is excluded from forecast evaluation to allow the forecasts time to `learn' sufficiently about the strengths and weaknesses of the teams in a given season. All leagues include data up to and including the end of the 2018/19 season. \par

\begin{table}[!htb]
\begin{center}
\begin{tabular}{|l|rrr|}
\hline
League & No. matches & Match data available & Excluding burn-in \\ 
\hline
Belgian Jupiler League & 5090 & 480 & 384 \\
English Premier League & 9120 & 7220 & 5759 \\ 
English Championship & 13248 & 10484 & 8641 \\
English League One & 13223 & 10460 & 8608 \\  
English League Two & 13223 & 10459 & 8613 \\
English National League & 7040 & 5352 & 4642 \\  
French Ligue 1 & 8718 & 4907 & 4126 \\
French Ligue 2 & 7220 & 760 & 639 \\
German Bundesliga & 7316 & 5480 & 3502 \\ 
German 2.Bundesliga & 5670 & 1057 & 753 \\
Greek Super League & 6470 & 477 & 381 \\
Italian Serie A & 8424 & 5275 & 4439 \\
Italian Serie B & 8502 & 803 & 680 \\
Netherlands Eredivisie & 5814 & 612 & 504 \\
Portuguese Primeira Liga & 5286 & 612 & 504 \\
Scottish Premier League & 5208 & 4305 & 3427 \\
Scottish Championship & 3334 & 524 & 297 \\
Scottish League One & 3335 & 527 & 298 \\
Scottish League Two & 3328 & 525 & 297 \\
Spanish Primera Liga & 8330 & 5290 & 4449 \\
Spanish Segunda Division & 8757 & 903 & 771 \\
Turkish Super lig & 5779 & 612 & 504 \\
\hline
Total & 162435 & 77124 & 62218 \\
\hline
\end{tabular}
\caption{Data used in this paper.}
\label{table:Leagues_available}
\end{center}
\end{table}

\section{A model for predicting shot success} \label{section:model}
We propose a simple model for predicting the probability of a football team scoring from a shot at goal.  We are primarily interested in estimating the probability pre-match and therefore we do not take into account any specific information about the location or nature of a shot.  In short, in a match between two teams, we ask the question `If a particular team takes a shot, what is the probability that they score as a result?' \par

Consider a football league with $T$ teams that play each other over the course of a season.  Let $a_{1},...,a_{T}$ and $d_{1},...,d_{T}$ be attacking and defensive ratings respectively for each team.  In a match with the $i$-th team at home to the $j$-th team, the forecast probability of a home goal given a home shot is given by
\begin{equation}
p(G_{h})=\frac{1}{1+exp\{-m_{h}\}}
\end{equation}
where $m_{h}=c+h+\frac{1}{2}(a_{i}+d_{j})$.  Here, $c$ is a constant parameter and $h$ a parameter that allows for home advantage (if any).  \par

The forecast probability of an away goal given an away shot is given by
\begin{equation}
p(G_{a})=\frac{1}{1+exp\{-m_{a}\}}
\end{equation}
where $m_{a}=c-h+\frac{1}{2}(a_{j}+d_{i})$. \par

Here, we have a total of $2T+2$ parameters to be estimated.  We take a maximum likelihood approach with a slight adjustment such that more recent matches are given a higher weight than those that were played longer ago.  To do this, we make use of the `half life' approach taken by \cite{ley2019ranking} in which the weighting placed on the $m$-th match is determined by 

\begin{equation} \label{eq:wtime}
w_{time,m}(x_m)=\left(\frac{1}{2}\right)^{\frac{x_m}{H}}, 
\end{equation}

where $x_m$ is the number of days since the $m$-th match was played and $H$ is the `half life', that is the number of days until the weighting halves.

The likelihood function, adjusted with the half life parameter, is given by
\begin{equation}
L=\prod_{m=1}^{M} \phi(p_m,O_m)^{w_{time,m}(x_m)}
\end{equation}
where 
\begin{equation} 
\phi(a,b)=\begin{cases}
   a & \text{if } b=1, \\
   1-a & \text {if } b = 0.
\end{cases}
\end{equation}

The model requires the simultaneous optimisation of $2T+2$ parameters. In the experiments performed in this paper, we use the `fmincon' function in Matlab and select the `interior point' algorithm which provides a compromise between speed and accuracy. We set the constraints $\sum_{i=1}^{T} a_{i} = 0$ and $\sum_{i=1}^{T} d_{i} = 0$ so that all of the ratings are distributed around zero.  All parameters are initialised to zero in the optimisation algorithm. \par

\subsection{Forecast skill and reliability}
If our forecast model of shot success described in section~\ref{section:model} is to be useful, it is important to show that the forecasts it produces are informative in terms of predicting the probability of scoring from a shot at goal. In this section, we evaluate the performance of the forecasts and examine the effect of the half life parameter. \par

To evaluate whether the forecasts are informative at all, we can investigate whether they outperform a very simple system in which forecasts consist of the historical shot success frequency over all past matches. If our forecasts are able to outperform this simple system, we have shown there is value in taking into account the strengths of the teams involved. \par

In weather forecasting, the simple forecasting system described above is often called the `climatology' and we adopt this terminology. The climatology is commonly used as a benchmark for the skill of a set of forecasts and if the forecasts cannot outperform the climatology, the forecast system is of little value (\cite{katz2005economic}). Formally, in our case, the climatological probability $p(G)$ of scoring given a shot at goal takes the form
\begin{equation} \label{eq:clim}
p_c=\frac{\sum_{m=1}^M G_{m}}{\sum_{m} S_{m}}
\end{equation}
where $G_m$ and $S_m$ are the total number of goals and shots respectively in the $m$-th match and $M$ is the number of past matches considered. \par

Probabilistic forecasts are best evaluated using scoring rules.  The Ignorance and Brier scores, described in appendix~\ref{section:scoring_rules}, are two examples of scoring rules that are suitable for evaluating binary probabilistic forecasts and we consider the skill according to both.  For context, in each case, the score is given with that of the climatology subtracted such that, if the relative score is negative, the forecasts can be considered to be skillful. \par

\begin{figure}[!htb]
    \centering
    \includegraphics[scale=0.9]{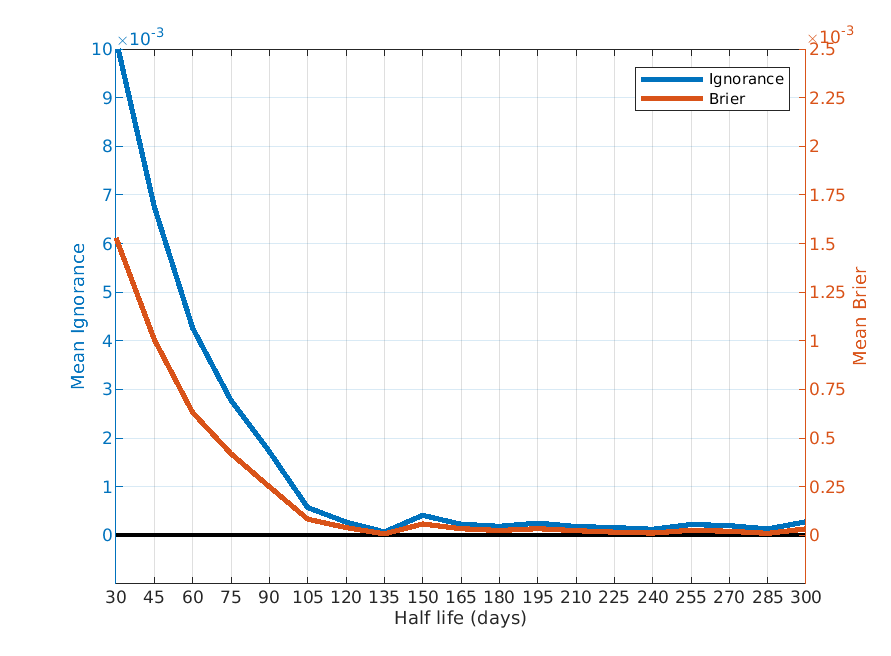}
    \caption{Mean Ignorance (blue line, left axis) and Brier (red line, right axis) scores for forecasts of the probability of scoring from a shot at goal, given relative to the climatology as a function of the half life parameter.}
    \label{figure:ign_RPS_function_halflife_just_unblended}
\end{figure}

The mean Ignorance and Brier scores of the forecasts relative to the climatology are shown as a function of the half life parameter in figure~\ref{figure:ign_RPS_function_halflife_just_unblended}. Here, the forecast skill under both scoring rules is positive for all values of the half life parameter implying that the forecasts do not outperform the climatology, on average. \par

To investigate why the forecasts are unable to outperform the climatology, we can make use of reliability diagrams to attempt to diagnose whether there are any systematic biases.  Reliability diagrams are used to visualise the `reliability' of a set of forecasts, that is whether the observed frequencies are consistent with the forecast probabilities (\cite{brocker2007increasing}). The forecasts are divided into `bins' and the mean forecast probability within each bin is plotted against the relative frequency of the outcomes.  If the points are close to the diagonal, the forecasts are `reliable'. We make use of the approach taken by \cite{brocker2007increasing} in which `consistency bars' are added which provide a 95 percent interval for the relative frequency under the assumption that the forecasts are perfectly reliable (that is, the outcomes occur at the rate implied by the forecasts). \par

\begin{figure}[!htb]
    \centering
    \includegraphics[scale=0.75]{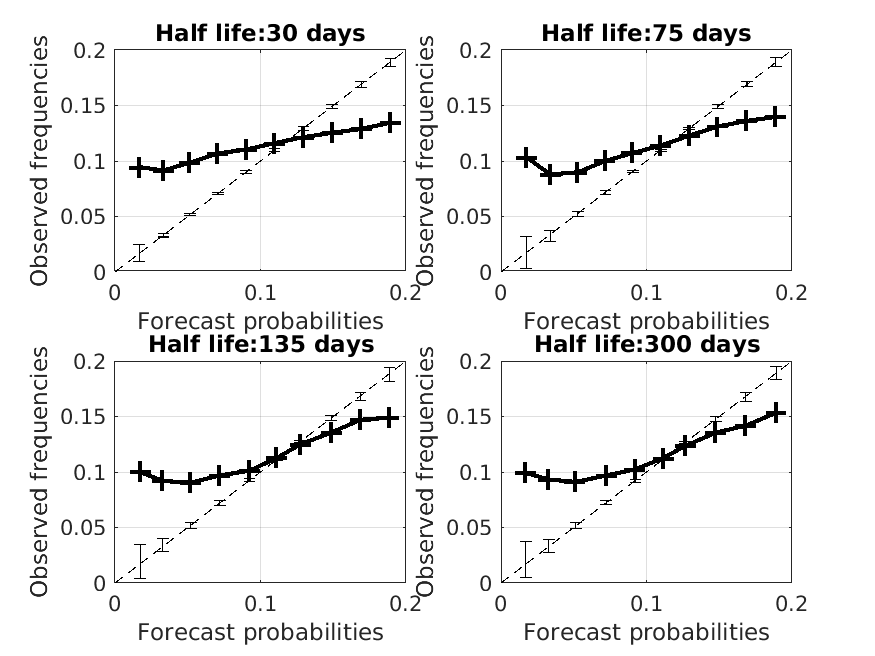}
    \caption{Reliability diagrams for forecasts of shot success for different values of the half life parameter.  The consistency bars show the region in which there is a 95 percent probability of the relative frequencies falling if the forecasts are perfectly reliable.}
    \label{figure:reliability_shotprobs}
\end{figure}

Reliability diagrams for different values of the half life parameter $H$ are shown in figure~\ref{figure:reliability_shotprobs}.  Here, in all cases, it is clear that the forecasts are overdispersed.  The highest forecast probabilities tend to correspond to far lower relative frequencies than would be expected if they were reliable, whilst the lowest forecast probabilities tend to correspond to much higher relative frequencies than expected.  To understand why we see the above pattern, it is useful to recall how the forecasts are formed.  The model assigns attacking and defensive parameters to each team as well as constant and home advantage parameters.  This means that a large number of parameters are required to be optimised simultaneously and this risks overfitting, in which the model does not generalise well out of sample.  For example, suppose a team happens to score with a large proportion of its shots in recent matches.  This will be reflected in their rating but may be unsustainable in the longer term, leading to an overestimate of the probability of scoring from a shot.  Conversely, a team that happens to have scored from a low proportion of its shots may have its probability of scoring in future matches underestimated. \par

To attempt to deal with overfitting, we adjust the forecasts using two different approaches.  In the first, we attempt to calibrate the forecasts using Platt Scaling, a simple approach in which the original forecast is used as an input to a logistic regression with a `calibrated' forecast as the output (\cite{platt1999probabilistic}).  The adjusted forecast $\tilde{p}$ is therefore given by 
\begin{equation}
\tilde{p}=\frac{1}{1+exp(A+bp)},
\end{equation}
where $p$ is the original forecast and $A$ and $b$ are parameters to be optimised over past forecasts and outcomes.  We use Maximum Likelihood to optimise the parameters over all available past forecasts. \par

Our second approach is `Blending' (\cite{brocker2008ensemble}).  Under this approach, the adjusted forecasts are a weighted average of the original forecast and the climatology (that is the historical average, see equation~\ref{eq:clim}).  Formally, the blended forecast is given by
\begin{equation}
\tilde{p}=\alpha p +(1-\alpha) p_c
\end{equation}
where $p$ is the original forecast, $p_c$ is the climatology and $\alpha$ is a parameter to be estimated.  Parameter estimation is done by minimising the mean ignorance score over all past forecasts (note this is equivalent to the Maximum Likelihood approach used in Platt Scaling). \par

\begin{figure}[!htb]
    \centering
    \includegraphics[scale=0.9]{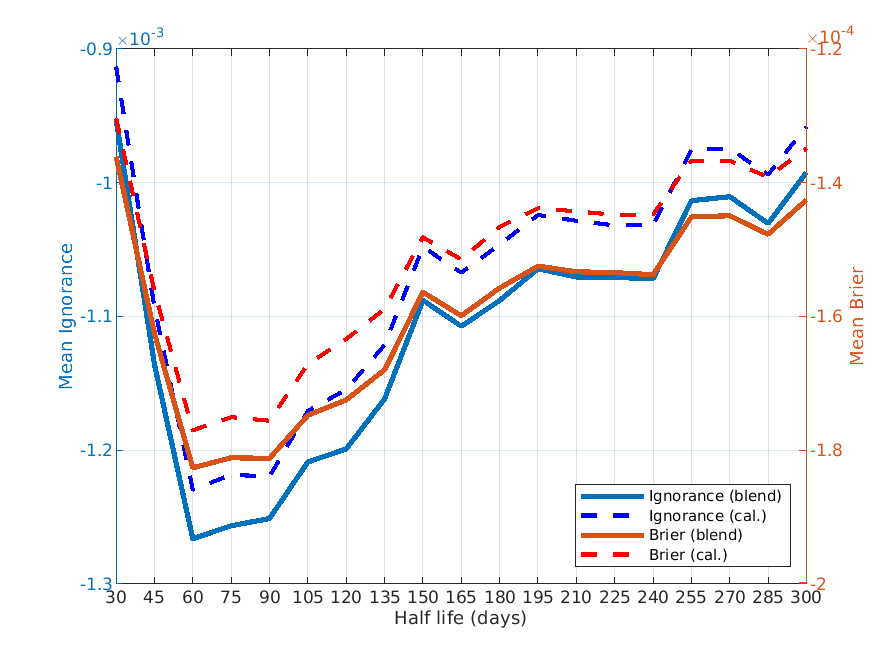}
    \caption{Mean Ignorance (blue line, left axis) and Brier (red line, right axis) scores for Blended (solid lines) and Platt Scaled (dashed lines) forecasts of the probability of scoring from any shot, given relative to the climatology, as a function of the half life parameter.}
    \label{figure:ign_RPS_function_halflife_just_blended_and_cal}
\end{figure}

The mean Ignorance and Brier scores (both shown relative to that of the climatology) of the Platt scaled and blended forecasts are shown in figure~\ref{figure:ign_RPS_function_halflife_just_blended_and_cal} (note the change in scale on the $y$ axis from figure~\ref{figure:ign_RPS_function_halflife_just_unblended}). Here, unlike the original forecasts, both the Platt scaled and blended forecasts produce negative mean Ignorance and Brier scores and are therefore able to outperform the climatology, demonstrating forecast skill. \par

\begin{figure}[!htb]
    \centering
    \includegraphics[scale=0.65]{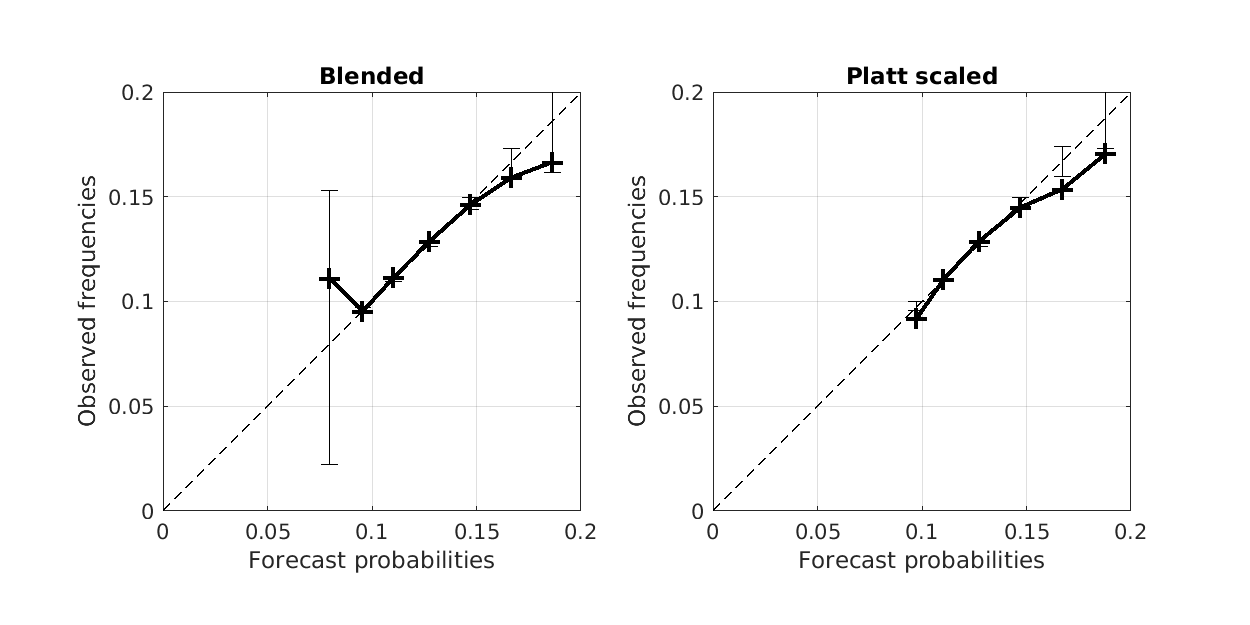}
    \caption{Reliability diagrams for forecasts of shot success adjusted using blending (left) and Platt Scaling (right) with a half life parameter of 60 days.  The consistency bars show the region in which there is a 95 percent probability of the relative frequencies falling if the forecasts are perfectly reliable.}
    \label{figure:reliability_shotprobs_cal_blend_side_by_side}
\end{figure}

It is clear that the choice of the half life parameter is crucial in determining the skill of the forecasts.  If it is too high, matches that were played a long time ago and have low relevance to the current time are given too much weight.  If it is too low, recent matches are given too little weight and the ratings assigned to each team are not robust. Here, under both scores and both approaches, the optimal half life parameter (out of those considered) is 60 days indicating that relatively recent matches play the biggest role in determining the probability of scoring.  It is also clear that the blending approach consistently outperforms Platt Scaling.  Reliability diagrams for the forecasts produced under Blending and Platt Scaling with a half life parameter of 60 days are shown in figure~\ref{figure:reliability_shotprobs_cal_blend_side_by_side}.  Under both approaches, it is clear that the effect is to moderate the forecasts by moving them closer to the climatology, creating improved reliability and skill. \par

In summary, the results here show that, when combined with Platt Scaling or Blending, our model is able to make skillful predictions of the probability of scoring from a given shot.  Having shown that we are able to construct skillful shot success forecasts, we now investigate whether they are effective in improving the skill of forecasts of match outcomes and whether the total number of goals in a match will exceed 2.5. \par

\section{Forecasting match outcomes and total goals} \label{section:predicting_match_outcomes}
In this section, we investigate whether our shot success model can be used alongside predictions of the number of shots to make informative probabilistic forecasts for (i) the outcomes of football matches (i.e. whether the match will end as a home win, draw or away win) and (ii) whether the total number of goals will exceed $2.5$ (henceforth `over/under 2.5 goal forecasts').  In each case, we assess both the forecast skill and the profitability when using the resulting forecasts alongside the two betting strategies defined in appendix~\ref{section:betting_strategy}.

Given a point prediction of the number of shots and the forecast probability of each of those shots being successful, we can obtain a point estimate for the number of goals scored by each team in a match by simply multiplying them together.  To predict the number of shots achieved by each team, we make use of the Generalised Attacking Performance (GAP) rating system proposed by \cite{wheatcroft2020} which has been shown to be a useful predictor variable for producing over/under 2.5 goal forecasts and forecasts of the match outcome (\cite{wheatcroft2019evaluating}).  The system is described in detail in appendix~\ref{section:GAP_ratings}.  Define a point prediction of the number of goals for the home team in a match to be 
\begin{equation} \label{equation:exp_goals_h}
E_{h}=\hat{S}_{h}P(G_{h})
\end{equation}
and, for the away team,
\begin{equation} \label{equation:exp_goals_a}
E_{a}=\hat{S}_{a}P(G_{a}),
\end{equation}
where $\hat{S}_{h}$ and $\hat{S}_{a}$ are the predicted number of shots for the home and away teams respectively, and $P(G_{h})$ and $P(G_{a})$ are the predicted probabilities that the home or away team will score given they have taken a shot at goal.  Note that $E_{h}$ and $E_{a}$ will usually not be integer values and represent a prediction of the `expected' number of goals achieved by each team.   \par 

For comparison, we can define a point prediction for the number of goals adjusted with the climatological probability such that
\begin{equation} \label{equation:exp_goals_h_clim}
C_{h}=\hat{S}_{h}p_c
\end{equation}
and 
\begin{equation} \label{equation:exp_goals_a_clim}
C_{a}=\hat{S}_{a}p_c 
\end{equation}
for the home and away teams respectively where $p_c$ is the climatological probability of shot success (i.e. the probability of a team scoring from a shot regardless of ability). \par

We make use of ordered logistic regression to map predictor variables into forecast probabilities for the match outcome. The ordered logistic regression model is chosen because the outcomes of football matches can be considered `ordered'.  In a sense, a home win and a draw are `closer together' than a home win and an away win and this is reflected in the parametrisation of the model. The ordered regression model allows $K$ predictor variables to be mapped into forecast probabilities.  A sensible choice of predictor variable for the match outcome is the difference in the predicted number of goals scored by each team defined by 
\begin{equation} \label{equation:exp_goals}
V=E_{h}-E_{a}.
\end{equation}

We use logistic regression to build probabilistic forecasts of whether the total number of goals in a match will exceed $2.5$. Since this is a binary event, logistic regression is a suitable model for mapping predictor variables to probabilities. Since we are interested in the total number of goals scored in a match, we use as a predictor variable the sum of the predicted number of goals scored by the home and away teams. The predictor variable is therefore 
\begin{equation} \label{equation:total_goals}
V=E_{h}+E_{a}.
\end{equation}

For our model of shot success to be effective in terms of predicting the match outcome and whether the number of goals will exceed 2.5, our predictor variables should be more informative than when $E_{h}$ and $E_{a}$ are replaced with $C_{h}$ and $C_{a}$, that is the case in which the probability of shot success is taken to be that of the climatology.  This comparison is the main focus of our experiment. \par
  
In addition to the predictor variables specified above, we consider the use of odds-implied probabilities as additional predictor variables. The rationale of this is that we may be able to `augment' the substantial information in the odds with additional information to provide more skillful forecasts. \par

\subsection{Experimental design}
We make use of the data described in section~\ref{section:data} to produce probabilistic forecasts both for the match outcome and for whether the total number of goals in a match will exceed $2.5$.  We do this for each match in which both shot data and the relevant odds are available.  This means we have a total of 62218 forecasts of the match outcome and 53447 over/under 2.5 goal forecasts. We produce two sets of forecasts in each case.  First we include in the model only our chosen predictor variable based on the predicted number of goals. Second, we include an odds-implied probability as an additional variable.  In the match outcome case, this is the odds-implied probability of a home win and, in the total goals case, the odds-implied probability that the total number of goals will exceed $2.5$. \par

In all cases, the forecasts for each match are constructed using regression parameters fitted with least squares estimation on all available matches in all leagues up to the day before the match is played.  In order to allow the forecasts to have sufficiently learned about the quality of the teams, we follow the approach of \cite{wheatcroft2020} and allow a `burn-in' period, thus excluding from calculations of forecast skill and profit the first six matches of the season for each team. \par

Since we are primarily interested in the potential value added by our shot success model, our comparison of interest is between the forecasts produced using as predictor variables the predicted number of goals calculated using our shot success model (that is formed using equations~(\ref{equation:exp_goals_h}) and~(\ref{equation:exp_goals_a})), and those produced using the climatological shot success probability defined in equation~(\ref{eq:clim}).  The latter case includes no information about the strength of the teams and therefore the extent to which it is able to be outperformed by our shot success model demonstrates its value to the forecasts.  We therefore present the skill of the forecasts formed using our shot success forecasts `relative' to those formed using the climatological probability of shot success.  This is done by subtracting the skill of the latter from the former such that negative values imply better relative skill. \par

We also compare the betting performance under the Level Stakes and Kelly betting strategies described in section~\ref{section:betting_strategy}. To calculate the overall profit, we use the maximum odds available from the BetBrain odds-comparison website, which are included in the `football-data' data set. \par

\subsection{Forecasts of the match outcome}
\begin{figure}[!htb]
    \centering
    \includegraphics[scale=0.9]{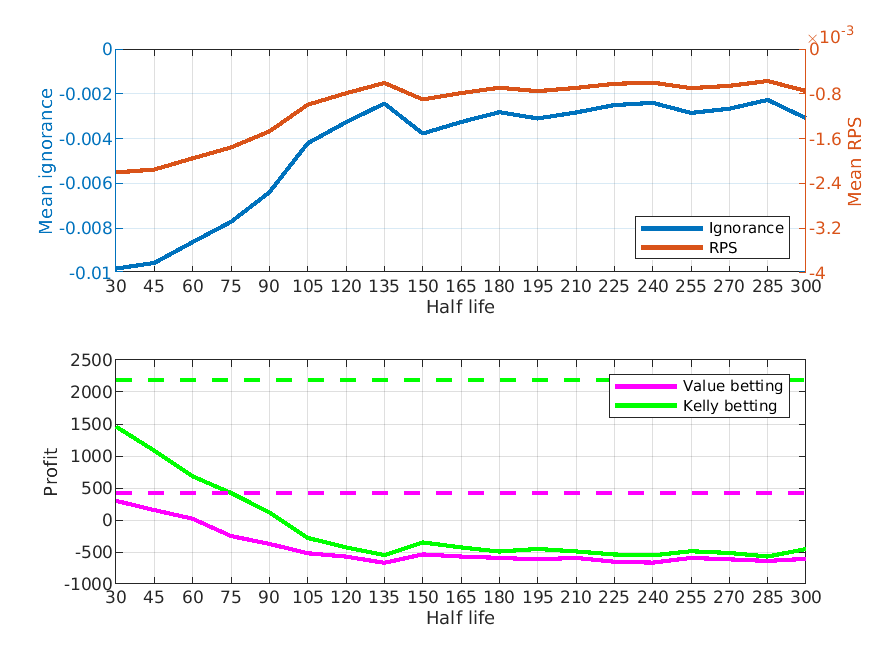}
    \caption{Top panel: Mean ignorance (blue line, left axis) and RPS (red line, left axis) for forecasts of the match outcome as a function of the half life parameter when the odds-implied probability is not included as an additional predictor variable.  Both scores are given relative to that of the case in which the predicted number of goals is calculated using the climatological probability of scoring.  Lower panel: Overall profit from the Level Stakes (magenta) and Kelly (green) strategies as a function of half life.  The dashed horizontal lines show the overall profit when the predicted number of goals is calculated using the climatological probability of scoring.}
    \label{figure:skill_and_profit_func_halflife_without_odds}
\end{figure}

We begin by considering forecasts of the match outcome. The mean relative Ignorance and Ranked Probability Scores for the case in which the odds-implied probability is not included as an additional predictor variable are shown as a function of the half life parameter $H$ in the top panel of figure~\ref{figure:skill_and_profit_func_halflife_without_odds}.  As described above, in both cases, the skill is given relative to (i.e is subtracted from) that of forecasts formed using the predicted number of goals adjusted using the climatological probability of shot success.  Since both the mean relative ignorance and RPS are negative, the shot success forecasts are shown to add skill to the match outcome forecasts for all considered values of the half life. \par

The overall profit under the Level Stakes (magenta) and Kelly (green) betting strategies are shown in the lower panel.  The dashed line shows the overall profit for the case in which the predicted number of goals is calculated using the climatological probability of the rate of shot success.  Interestingly, despite the fact that our shot success model improves forecast skill, the overall profit is slightly decreased and there is therefore no evidence of improved gambling performance under either strategy.  Both forecast skill and the overall profit are optimised by setting the half life parameter to 30 days, implying that shot success in relatively recent matches is the most informative in terms of the match outcome. \par

\begin{figure}[!htb]
    \centering
    \includegraphics[scale=0.9]{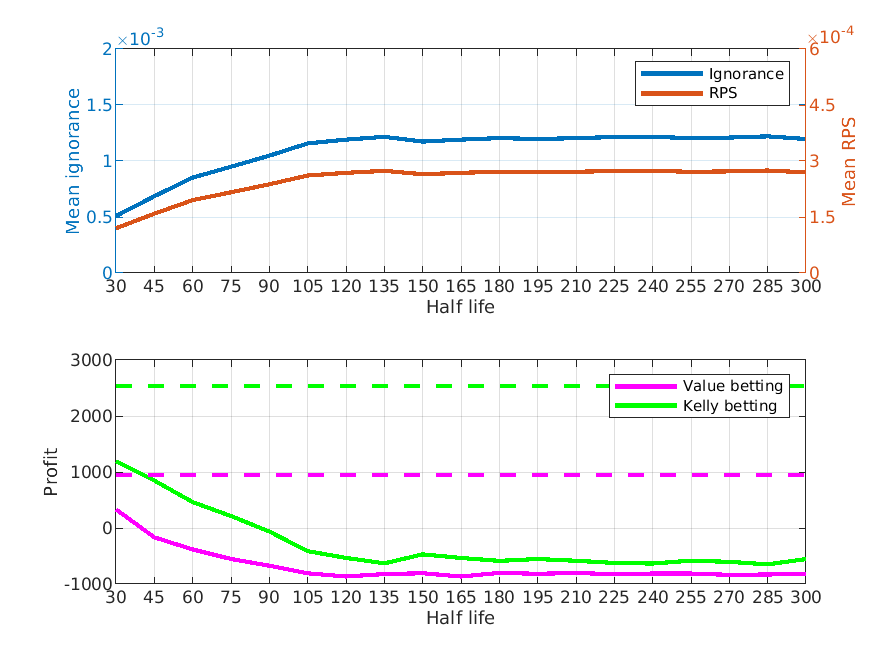}
    \caption{Top panel: Mean ignorance (blue line, left axis) and RPS (red line, left axis) for forecasts of the match outcome as a function of the half life parameter when the odds-implied probability is included as an additional predictor variable.  Both scores are given relative to that of the case in which the predicted number of goals is calculated using the climatological probability of scoring.  Lower panel: Overall profit from the Level Stakes (magenta) and Kelly (green) strategies as a function of half life.  The dashed horizontal lines show the overall profit when the predicted number of goals is calculated using the climatological probability of scoring.}
    \label{figure:skill_and_profit_func_halflife_with_odds}
\end{figure}

Figure~\ref{figure:skill_and_profit_func_halflife_with_odds} is the same as figure~\ref{figure:skill_and_profit_func_halflife_without_odds} but for the case in which the odds-implied probability of a home win is included as an additional predictor variable. Here, both the relative ignorance and RPS are positive, implying that our model of shot success is counterproductive. Similarly, there is a reduction in profit under both betting strategies.  We can provide a speculative answer as to why this is the case.  Betting odds are complex and reflect a great deal of information brought together by participants in the market.  We suggest that differences in the probability of shot success are efficiently reflected in the odds (punters may account for efficient goal scorers/goalkeepers etc.) and therefore, by including this information, there is an element of double counting which negatively impacts the forecasts.  It is worth noting that finding information that can `augment' the information in the betting odds is a much more difficult task than finding information to produce forecasts from scratch.  We discuss this further in section~\ref{results_summary}. \par

\subsection{Over/under 2.5 goal forecasts}
\begin{figure}[!htb]
    \centering
    \includegraphics[scale=0.9]{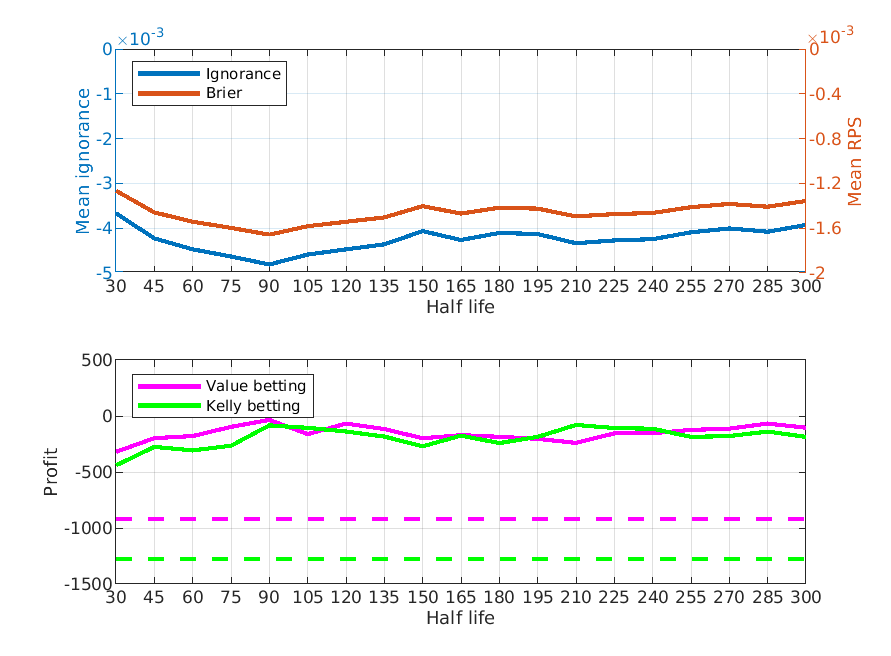}
    \caption{Top panel: Mean ignorance (blue line, left axis) and Brier score (red line, left axis) for the over/under 2.5 goal forecasts as a function of the half life parameter when the odds-implied probability is not included as a predictor variable.  Both scores are given relative to that of the case in which the predicted number of goals is calculated using the climatological probability of scoring. Lower panel: Overall profit from the Level Stakes (magenta) and Kelly (green) strategies as a function of the half life parameter.  The dashed horizontal lines show the overall profit when the predicted number of goals is calculated using the climatological probability of shot success.}
    \label{figure:skill_and_profit_func_halflife_without_odds_OU}
\end{figure}

We now turn to the over/under 2.5 goal forecasts. The results for the case in which the odds-implied probability is not included as an additional predictor variable are shown in figure~\ref{figure:skill_and_profit_func_halflife_without_odds_OU}. Similarly to the forecasts of the match outcome, here, the top panel shows the mean Ignorance and Brier scores given relative to the case in which the forecasts are formed using the predicted number of goals produced using the climatological probability of the rate of shot success. Since both relative scores are negative, our shot success model is able to increase the skill of the forecasts. \par

The overall profit achieved using the Kelly and Level Stakes betting strategies is shown in the lower panel of figure~\ref{figure:skill_and_profit_func_halflife_without_odds_OU}. Here, as before, the solid lines show the overall profit for the case in which the predicted number of goals are produced using our shot success model and the dashed lines the case in which the climatological rate of shot success is used. Here, there is a major improvement in the gambling return from using our model of shot success, although the profit is still slightly negative for all values of the half life parameter. The optimal half life parameter of 90 days is slightly longer than for the match outcome forecasts but this still suggests that relatively recent matches are most relevant. \par

\begin{figure}[!htb]
    \centering
    \includegraphics[scale=0.9]{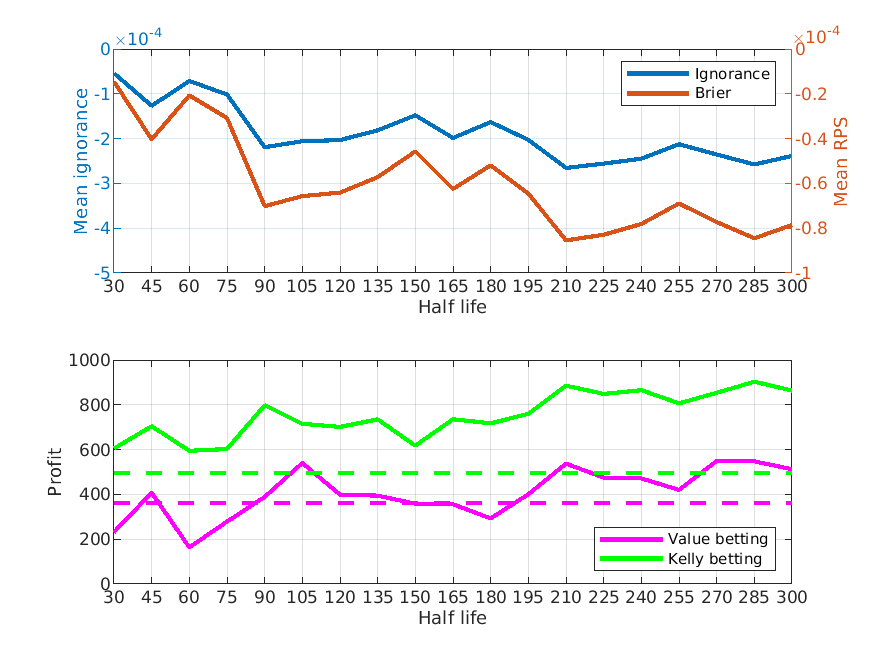}
    \caption{Top panel: Mean ignorance (blue line, left axis) and Brier score (red line, left axis) for the over/under 2.5 goal forecasts as a function of the half life parameter when the odds-implied probability is included as a predictor variable.  Both scores are given relative to that of the case in which the predicted number of goals is calculated using the climatological probability of scoring. Lower panel: Overall profit from the Level Stakes (magenta) and Kelly (green) strategies as a function of the half life parameter.  The dashed horizontal lines show the overall profit when the predicted number of goals is calculated using the climatological probability of shot success.}
    \label{figure:skill_and_profit_func_halflife_with_odds_OU}
\end{figure}

Figure~\ref{figure:skill_and_profit_func_halflife_with_odds_OU} shows the same results as figure~\ref{figure:skill_and_profit_func_halflife_without_odds_OU} but for the case in which the odds-implied probability is included as an additional predictor variable. Here, the relative skill under both the Ignorance and Brier scores is negative, implying that our shot success model increases the skill of the over/under 2.5 goal forecasts.  For most values of the half life parameter, there is also an increase in profit under both strategies.  This is a very different result to the match outcome case in which we were unable to improve the forecasts using our model of shot success. Interestingly, the most effective choice of half life parameter is 300 days, suggesting that shot success over a longer period of time is relevant here. \par

\subsection{Summary} \label{results_summary}
The results above demonstrate that our model for predicting shot success can improve the skill of shot-based forecasts for both match outcomes and for whether the total number of goals will exceed 2.5.  For the case in which the odds-implied probability is not included in the forecasts, gains in forecast skill are demonstrated for both sets of forecasts.  For the case in which the odds-implied probability is included, the results are more mixed with an improvement in the skill of over/under 2.5 goal forecasts and a reduction in the skill of forecasts of the match outcome. \par

It is worth noting the philosophical difference between forecasts formed with and without the odds-implied probability included as an additional predictor variable.  In the latter case, we are building forecasts effectively from scratch and therefore it should be relatively straightforward to find information that adds to the skill.  We know that we are able to build skillful forecasts using predicted match statistics and, logically, if we can incorporate skillful forecasts of the rate of shot success, we should be able to improve the forecasts and this has proven to be the case.  In the former case, we have a very different situation.  Betting odds are generally considered to be highly informative reflections of the underlying probability of an outcome (though there are a number of known biases), taking into account a wide range of factors. Finding information that can `augment' this information is therefore a much more difficult task.  Further, there is likely to be a complicated relationship between our forecasts of the rate of shot success and the extent to which this information is reflected in the odds. The fact that we are able to improve the over/under 2.5 goal forecasts but not the match outcome forecasts is testament to this complex relationship.  \par

It is less clear whether the general improvement in skill achieved from the forecasts of the rate of shot success leads to increased gambling profit.  This probably reflects the complex relationship between forecast probabilities and gambling returns.  For both the Level Stakes and Kelly strategies, gambling success is dependent on finding bets that offer a positive expected return.  Success at doing this, however, will not necessarily increase with forecast skill. Consider the Level Stakes case.  Here, a bet is taken if the forecast probability is higher than the odds-implied probability and the forecast therefore implies that there is value. The success of the strategy is dependent on the forecasts successfully identifying bets in which there is genuine value.  If improvements in skill are largely seen in forecasts in which the decision as to whether to bet or not is unchanged and reductions in skill are in `borderline' cases, it is easy to see how the profit may fall with increased average forecast skill.  This is not a criticism of the approach of using scoring rules to evaluate forecasts but rather a demonstration of the difference between forecast skill and the utility of using the forecasts for a particular decision process. \par

\section{Discussion} \label{section:discussion}
In this paper, we have presented a model for predicting the probability of a football team scoring from a shot at goal.  Whilst the model suffers from overfitting, we are able to calibrate the forecasts to produce good forecast skill.  We have also demonstrated that the model of shot success can be used alongside predictions of the number of shots achieved by each team to provide improved skill for both match outcome and over/under 2.5 goal forecasts. \par

Whilst our shot success model has been shown to be able to produce improved forecast skill, there is also an economic interpretation of the results.  The experiments we have conducted were partly inspired by the results shown in \cite{wheatcroft2020} and \cite{wheatcroft2019evaluating} that showed that predicted match statistics, formed using GAP ratings, can provide forecast skill beyond that reflected in the odds.  We have built on this and shown that, for over/under 2.5 goal forecasts, we can provide further improvement using our forecasts of the rate of shot success.  As described in the aforementioned papers, the fact that predicted match statistics can improve a set of forecasts has implications for the efficiency of the betting markets, implying that the market does not efficiently account for this information.  The results in this paper build on that and suggest that the over/under 2.5 goal market does not adequately account for the probability of scoring from a shot.  We do not have evidence that this is the case for the match outcome market, however. \par 

In our opinion, there is potential value in the model beyond those applications demonstrated here.  It is, of course, desirable for a team to score with a relatively high proportion of shots, since doing so would result in more goals and better match results.  Similarly, it is desirable to concede from a relatively small proportion of shots.  A manager looking to improve their team's results may be interested both in the quality of their players' shot conversion and the ability of their defence to prevent the opposition from converting their shots.  However, simply looking at observed rates of shot conversion in recent matches would likely not give a robust estimate of their skill in converting shots to goals.  Our shot success forecasts are a potentially useful alternative to looking at observed numbers because they provide a more robust measure of the skill of each team since the half life and blending parameters have been chosen with respect to objective forecast skill.  This objectivity allows some of the inevitable biases of the manager to be removed when assessing the performance of their team. \par

Another interesting question regards the value of combining the model presented here with expected goals methodologies.  The idea behind expected goals is that the location and nature of each shot is used to provide an estimate of the probability of a shot ending with a goal.  The sum of the probabilities assigned to the shots in a match can then be interpreted as a measure of the number of goals a team would be `expected' to score, given the shots it has taken.  Importantly, expected goals typically don't take into account the relative abilities of the teams or players.  Conversely, it is important to note that, under our model, the nature of a shot is not taken into account.  This is potentially important because the location from which shots are taken have a big impact on the probability of scoring and some teams may be more likely to take shots from locations in which it is difficult to score, reducing their shot conversion rate.  In order to determine whether a rate of shot conversion is due to the nature of the shots or poor shooting ability, one could compare the probability of scoring from each shot under the expected goals methodology with the forecast probability of scoring under our model (which, unlike expected goals, takes into account the ability of the teams). If the latter is typically higher than the former, one might conclude that a team's shooting ability is high. \par

In conclusion, it is becoming increasingly clear that forecasts based on the number of shots at goal have great value in predicting the outcomes of football matches.  An obvious weakness of this approach is that the ability of the two teams involved is not taken into account.  This paper provides a potential solution to that weakness. \par 

\appendix

\section{GAP Rating System} \label{section:GAP_ratings}
The Generalised Attacking Performance (GAP) rating system was introduced by \cite{wheatcroft2020} and is a rating system designed to assess the attacking and defensive strength of sports teams with relation to some defined measure of attacking performance.  In this paper, we are interested in the number of shots taken by each team in football. For the chosen measure of attacking performance (in our case, shots), each team is given a separate attacking and defensive rating for its home and away matches such that it has four ratings in total.  An attacking GAP rating is interpreted as an estimate of the number of defined attacking plays the team can be expected to achieve against an average team in the league. Its defensive rating can be interpreted as an estimate of the number of attacking plays it can be expected to concede against an average team. A team's ratings are updated each time it plays a match. The GAP ratings of the $i$-th team for its $k$-th match in a league are denoted as follows:
\begin{itemize}
\item $H_{i,k}^{a}$ - Home attacking GAP rating of the $i$-th team in a league after $k$ matches.
\item $H_{i,k}^{d}$ - Home defensive GAP rating of the $i$-th team in a league after $k$ matches.
\item $A_{i,k}^{a}$ - Away attacking GAP rating of the $i$-th team in a league after $k$ matches.
\item $A_{i,k}^{d}$ - Away defensive GAP rating of the $i$-th team in a league after $k$ matches.
\end{itemize}
For a match involving the $i$-th team at home to the $j$-th team, the predicted numbers of shots achieved by the home and away teams respectively are given by
\begin{equation}
\hat{S}_{h}=\frac{H_{i}^{a}+A_{j}^{d}}{2}
\hat{S}_{a}=\frac{A_{j}^{a}+H_{i}^{d}}{2}.
\end{equation}
In other words, each team's predicted number of attacking plays is given by the average of its attacking rating and its opponent's defensive rating. Updates are performed as follows. Consider a match in which the $i$-th team in the league is at home to the $j$-th team and in which the $i$-th team has played $k_{1}$ previous matches and the $j$-th team $k_{2}$. Let $S_{i,k_{1}}$ and $S_{j,k_{2}}$ be the number of defined attacking plays by teams $i$ and $j$ in the match. The GAP ratings for the $i$-th team (the home team) are updated as follows:
\begin{equation} 
\begin{split}
H_{i,k_{1}+1}^{a} & = \max(H_{i,k_{1}}^{a}+\lambda\phi_{1}(S_{i,k_{1}}-(H_{i,k_{1}}^{a}+A_{j,k_{2}}^{d})/2),0), \\
A_{i,k_{1}+1}^{a} & = \max(A_{i,k_{1}}^{a}+\lambda(1-\phi_{1})(S_{i,k_{1}}-(H_{i,k_{1}}^{a}+A_{j,k_{2}}^{d})/2),0), \\
H_{i,k_{1}+1}^{d} & = \max(H_{i,k_{1}}^{d}+\lambda\phi_{1}(S_{j,k_{2}}-(A_{j,k_{2}}^{a}+H_{i,k_{1}}^{d})/2),0), \\
A_{i,k_{1}+1}^{d} & = \max(A_{i,k_{1}}^{d}+\lambda(1-\phi_{1})(S_{j,k_{2}}-(A_{j,k_{2}}^{a}+H_{i,k_{1}}^{d})/2),0) \\
\end{split}
\end{equation}
The GAP ratings for the $j$-th team (the away team) are updated according to:
\begin{equation} 
\begin{split}
A_{j,k_{2}+1}^{a} & = \max(A_{j,k_{2}}^{a}+\lambda\phi_{2}(S_{j,k_{2}}-(A_{j}^{a}+H_{i}^{d})/2),0), \\
H_{j,k_{2}+1}^{a} & = \max(H_{j,k_{2}}^{a}+\lambda(1-\phi_{2})(S_{j,k_{2}}-(A_{j}^{a}+H_{i}^{d})/2),0), \\
A_{j,k_{2}+1}^{d} & = \max(A_{j,k_{2}}^{d}+\lambda\phi_{2}(S_{i,k_{1}}-(H_{i}^{a}+A_{j}^{d})/2),0), \\
H_{j,k_{2}+1}^{d} & = \max(H_{j,k_{2}}^{d}+\lambda(1-\phi_{2})(S_{i,k_{1}}-(H_{i}^{a}+A_{j}^{d})/2),0), \\
\end{split}
\end{equation}
where $\lambda>0$, $0<\phi_{1}<1$ and $0<\phi_{2}<1$ are parameters to be estimated. The role of $\lambda$ is to determine the overall influence of a match on the ratings of each team.  The parameters $\phi_{1}$ and $\phi_{2}$ determine the impact of a home match on a team's away ratings and of an away match on a team's home ratings respectively.  After a given match, a home team is said to have outperformed expectations in an attacking sense if its attacking performance is higher than its predicted performance.  In this case, its attacking ratings are increased, whilst its ratings are decreased if it underperforms expectations. \par

GAP ratings are determined by three parameters which, as in \cite{wheatcroft2020}, are optimised using least-squares minimisation, with the aim of minimising the mean absolute error between the estimated and observed number of attacking plays. The function to be minimised is therefore
\begin{equation}
f=\frac{1}{N} \sum_{m=1}^{N} |S_{h,m}-\hat{S}_{h,m}|+|S_{a,m}-\hat{S}_{a,m}|
\end{equation}
where, for the $m$-th match, $S_{h,m}$ and $S_{a,m}$ are the observed numbers of attacking plays for the home and away team respectively and $\hat{S}_{h,m}$ and $\hat{S}_{a,m}$ are the predicted numbers from the GAP rating system. As in \cite{wheatcroft2020}, optimisation is performed using the fminsearch function in Matlab which implements the Nelder-Mead simplex algorithm. \par

\section{Scoring rules} \label{section:scoring_rules}
In this paper, we construct probabilistic forecasts for (i) the probability of scoring from a shot, (ii) the outcomes of football matches and (iii) whether the total number of goals in a match will exceed 2.5.  We evaluate the forecasts using scoring rules.  A scoring rule is a function of a probabilistic forecast and corresponding outcome aimed at evaluating forecast performance.  We make use of three different scoring rules and these are defined below. \par

Let an event have $r$ possible outcomes and let $p_{j}$ be the forecast probability at position $j$ where the ordering of the positions is preserved and $\sum_{i}^{r} p_{i}$.  Let $y \in \{1,...,r\}$ be the outcome and define $o_{1},...,o_{r}$ such that
\begin{equation}
o_{j}=\begin{cases}
1 & \text{if } j = y\\
0 & \text{otherwise }
\end{cases}
\end{equation}

\noindent The Brier Score (\cite{brier1950verification}) is defined as  
\begin{equation}
\mathrm{Brier}=\sum_{i=1}^{r}(p_{i}-o_{i})^{2}.
\end{equation}
The Ranked Probability Score (RPS) is defined (\cite{epstein1969scoring}) as
\begin{equation}
\mathrm{RPS}=\sum_{i=1}^{r-1}\sum_{j=1}^{i}(p_{j}-o_{j})^{2}.
\end{equation}
The ignorance score (\cite{10.2307/2984087,roulston2002evaluating}) is defined as 
\begin{equation}
\mathrm{IGN}=-\log_{2}(p_y).
\end{equation}

There is much debate surrounding choices of scoring rules and this usually centres on whether they have certain desirable properties.  It is widely agreed that scores should be \emph{proper} which means that, in expectation, no imperfect forecast will outperform a forecast coinciding with the `true' probability distribution (\cite{brocker2007scoring}).  All three of the above scores are proper and therefore forecasters are incentivised to give a forecast reflecting their true belief. Note that the RPS is only suitable for `ordered' outcomes.  We make use of the Ignorance and Brier scores to evaluate the binary shot forecasts and forecasts of whether the total number of goals in a match will exceed 2.5. We use the Ignorance and Ranked Probability Scores to evaluate the match outcome forecasts.  For a discussion on the relative merits of the three scoring rules in terms of evaluating forecasts of football matches, see \cite{wheatcroft2019evaluating}. \par

\section{Betting Strategies} \label{section:betting_strategy}
In section~\ref{section:predicting_match_outcomes}, we assess the performance of match forecasts in terms of betting performance.  To do this, we make use of two betting strategies: a simple level stakes value betting strategy and a strategy based on the Kelly Criterion. Both strategies are described below and follow the terminology described in \cite{wheatcroft2020}. \par

In this paper, we use decimal betting odds in which the odds offered on an event is simply the number by which the gambler's stake is multiplied in the event of success. Therefore, if the decimal odds are $2$, a \pounds 10 bet on said event would result in a return of $2 \times \pounds 10 = \pounds 20$. Let $o_{i}$ be the odds offered on the $i$-th potential outcome.  The \emph{odds-implied probability} $r_{i}=\frac{1}{o_{i}}$ is simply the multiplicative inverse of the odds. \par

The \emph{Level stakes} strategy is a simple value betting strategy in which a unit bet is placed on the $i$-th possible outcome of an event if $\hat{p}_{i}>r_{i}$, where $\hat{p}_{i}$ and $r_{i}$ are the predicted and odds-implied probabilities, respectively.  The rationale here is that, if the forecast implies that the true probability is higher than the odds-implied probability, the bet offers `value', that is a positive expected profit. \par

The \emph{Kelly strategy} is based on the Kelly Criterion (\cite{kelly1956new}) and, like the Level stakes strategy, is based on the concept of `value'.  However, under this strategy, the stake is dependent on the difference between the forecast probability and the odds-implied probability. When there is a large discrepancy, a higher stake is made. Under the Kelly Criterion, the amount staked is proportional to one's wealth.  For a particular outcome, the proportion of wealth staked is
\begin{equation} \label{eq:kelly}
f_{i}=\max\left(\frac{o_{i}+\hat{p}_{i}-1}{o_{i}-1},0\right)
\end{equation}
where $\hat{p}_{i}$ is the estimated probability of the outcome and $o_{i}$ represents the decimal odds on offer. Here, we do not bet proportionally to wealth but, rather, ensure that the average stake is 1 such that both betting strategies are directly comparable.  We therefore set the stake for the $i$-th bet to $s_{i}=kf_{i}$ where $k$ is a normalising constant set such that $\frac{1}{m} \sum_{i=1}^{m} kf_{i}=1$, $f_{i}$ is calculated from equation~\ref{eq:kelly} and $m$ is the total number of bets placed. \par

\newpage
\bibliographystyle{agsm}
\bibliography{bibliography}
\end{document}